\documentstyle[12pt]{article}
\newcommand\be{\begin{equation}}
\newcommand\ee{\end{equation}}
\newcommand\bea{\begin{eqnarray}}
\newcommand\eea{\end{eqnarray}}
\newcommand\ket[1]{|#1\rangle}

\newcommand\braket[2]{\langle #1|#2\rangle}
\newcommand{\fatalpha}{{\bf \alpha \kern -0.44em \alpha}}
\newcommand{\fatsigma}{{\bf \sigma \kern -0.54em \sigma}}
\newcommand{\tpchi}{{\bf \chi \kern -0.35em \chi}}
\newcommand{\llambda}{{\bf \lambda \kern -0.45em \lambda}}

\renewcommand{\theequation}{\arabic{equation}}
\renewcommand{\theequation}{\thesection-\arabic{equation}}
\bibliography{plain}
\title{\bf Evaluation of effective resistances in pseudo-distance-regular resistor networks}\vspace{20mm}
\author{ M. A. Jafarizadeh$^{a,b,c}$
 \thanks{E-mail:jafarizadeh@tabrizu.ac.ir},
 R. Sufiani$^{a,b}$
 \thanks{E-mail:sofiani@tabrizu.ac.ir},
 S. Jafarizadeh$^{d}$
\\ $^a${\small Department of Theoretical Physics and Astrophysics,
University of Tabriz, Tabriz 51664, Iran.} \\ $^b${\small Institute
for Studies in Theoretical Physics and Mathematics, Tehran
19395-1795, Iran.} \\ $^c${\small Research Institute for Fundamental
Sciences, Tabriz 51664, Iran.}\\ $^d${\small Department of
Electrical and computer engineering, University of Tabriz, Tabriz
51664, Iran.}} \pagebreak

\vspace{20mm}
\begin{document}
\maketitle \vspace{15mm}
\newpage
\begin{abstract}
In Refs. \cite{jss1} and \cite{res2}, calculation of effective
resistances (two-point resistances) on distance-regular networks
was investigated, where in the first paper, the calculation was
based on the stratification of the network and Stieltjes function
associated with the network, whereas in the latter one a recursive
formula for effective resistances was given based on the
Christoffel-Darboux identity. In this paper, evaluation of
effective resistances on more general networks called
pseudo-distance-regular networks \cite{21} or $QD$ type networks
\cite{obata} is investigated, where we use the stratification of
these networks and show that the effective resistances between a
given node such as $\alpha$ and all of the nodes $\beta$ belonging
to the same stratum with respect to $\alpha$
($R_{\alpha\beta^{(m)}}$, $\beta$ belonging to the $m$-th stratum
with respect to the $\alpha$) are the same. Then, based on the
spectral techniques, an analytical formula for effective
resistances $R_{\alpha\beta^{(m)}}$ such that
$L^{-1}_{\alpha\alpha}=L^{-1}_{\beta\beta}$ (those nodes $\alpha$,
$\beta$ of the network such that the network is symmetric with
respect to them) is given in terms of the first and second
orthogonal polynomials associated with the network, where $L^{-1}$
is the pseudo-inverse of the Laplacian of the network. From the
fact that in distance-regular networks,
$L^{-1}_{\alpha\alpha}=L^{-1}_{\beta\beta}$ is satisfied for all
nodes $\alpha,\beta$ of the network, the effective resistances
$R_{\alpha\beta^{(m)}}$ for $m=1,2,...,d$ ($d$ is diameter of the
network which is the same as the number of strata) are calculated
directly, by using the given formula.

 {\bf Keywords:effective resistance,
pseudo-distance-regular networks, stratification, spectral
distribution}

{\bf PACs Index: 01.55.+b, 02.10.Yn }
\end{abstract}

\vspace{70mm}
\newpage
\section{Introduction}
The study of electric networks was formulated by Kirchhoff \cite{8}
more than 150 years ago as an instance of a linear analysis. Our
starting point is along the same line by considering the Laplacian
matrix associated with a network. The Laplacian is a matrix whose
off-diagonal entries are the conductances connecting pairs of nodes.
Just as in graph theory where everything about a graph is described
by its adjacency matrix (whose elements is 1 if two vertices are
connected and 0 otherwise), everything about an electric network is
described by its Laplacian. Recently, the authors have given a
method for calculation of the effective resistance on
distance-regular networks \cite{jss1}, where the calculation is
based on stratification introduced in \cite{js} and Stieltjes
transform of the spectral distribution (Stieltjes function)
associated with the network. Also, in Ref. \cite{jss1} it has been
shown that the resistances between a node $\alpha$ and all nodes
$\beta$ belonging to the same stratum with respect to the $\alpha$
($R_{\alpha\beta^{(i)}}$, $\beta$ belonging to the $i$-th stratum
with respect to the $\alpha$) are the same and the analytical
formulas have been given for two-point resistances
$R_{{\alpha\beta^{(i)}}},i=1,2,3$ in terms of the size of the
network and corresponding intersection array without any need to
know the spectrum of the pseudo inverse $L^{-1}$. In the next work
\cite{res2}, the authors have used the algebraic structure of
distance-regular networks (Bose-Mesner algebra) such as
stratification and spectral techniques specially the well known
Christoffel-Darboux identity \cite{tsc} from the theory of
orthogonal polynomials to give a recursive formula for calculation
of all of the resistance distances $R_{{\alpha\beta^{(i)}}}$,
$i=1,2,...,d$ ($d$ is diameter of the network which is the same as
the number of strata) on the network without any need to calculating
the spectrum of the pseudo inverse of the Laplacian of the network
denoted by $L^{-1}$. In this way they have shown that, the effective
resistance strictly increases by increasing the shortest path
distance defined on the network, i.e.,
$R_{{\alpha\beta^{(m+1)}}}-R_{{\alpha\beta^{(m)}}}>0$ for all
$m=1,2,...,d-1$. Here in this work, evaluation of effective
resistances on more general networks called pseudo-distance-regular
networks \cite{21} or $QD$ type networks \cite{obata} is
investigated, where we use the stratification of these networks and
show that the effective resistances between a given node such as
$\alpha$ and all of the nodes $\beta$ belonging to the same stratum
with respect to $\alpha$ are the same. Then, based on the spectral
techniques, an analytical formula for effective resistances
$R_{\alpha\beta^{(m)}}$ such that
$L^{-1}_{\alpha\alpha}=L^{-1}_{\beta\beta}$ (those nodes $\alpha$,
$\beta$ of the network such that the network is symmetric with
respect to them) is given in terms of the first and second
orthogonal polynomials associated with the network. Particularly,
due to the fact that, in distance-regular networks we have
$L^{-1}_{\alpha\alpha}=L^{-1}_{\beta\beta}$  for all nodes
$\alpha,\beta$ of the network, the effective resistances
$R_{\alpha\beta^{(m)}}$ for $m=1,2,...,d$ are calculated directly,
by using the given formula.

The organization of the paper is as follows. In section 2, we give
some preliminaries such as definitions related to graphs, their
adjacency matrices, stratification of the graphs,
pseudo-distance-regular graphs and spectral distribution associated
with the graphs. In section $3$, an analytical formula for
calculating the effective resistances in pseudo-distance-regular
graphs as resistor networks is given in terms of the orthogonal
polynomials of the first kind and second kind associated with the
networks. The paper is ended with a brief conclusion and an
appendix.
\section{Preliminaries}
In this section we give some preliminaries such as definitions
related to graphs, corresponding stratification,
pseudo-distance-regular graphs and spectral distribution techniques.
\subsection{Graph and its adjacency matrix}
In this section, we review the stratification of the graphs and the
notion of pseudo-distance-regularity.

A graph is a pair $\Gamma=(V,E)$, where $V$ is a non-empty set and
$E$ is a subset of $\{(x,y):x,y \in V,x\neq y\}$. Elements of $V$
and of $E$ are called a vertex and an edge, respectively. Two
vertices $x, y \in V$ are called adjacent if $(x,y)\in E$, and in
that case we write $x \sim y$. For a graph $\Gamma=(V,E)$, the
adjacency matrix $A$ is defined as
\begin{equation}\label{adj.}
\bigl(A)_{\alpha, \beta}\;=\;\cases{1 & if $\;\alpha\sim \beta$ \cr
0 & \mbox{otherwise}\cr}.
\end{equation}
Conversely, for a non-empty set $V$, a graph structure is uniquely
determined by such a matrix indexed by $V$.

The degree or valency of a vertex $x \in V$ is defined by
\begin{equation}\label{val.}
\kappa(x)=|\{y\in V: y\sim x\}|
\end{equation}
where, $|\cdot|$ denotes the cardinality. The graph is called
regular if the degree of all of the vertices be the same. In this
paper, we will assume that graphs under discussion are regular. A
finite sequence $x_0, x_1, . . . , x_n \in V$ is called a walk of
length $n$ (or of $n$ steps) if $x_{i-1}\sim  x_i$ for all $i= 1, 2,
. . . , n$. Let $l^2(V)$ denote the Hilbert space of $C$-valued
square-summable functions on $V$. With each $\beta\in V$ we
associate a vector $\ket{\beta}$ such that the $\beta$-th entry of
it is $1$ and all of the other entries of it are zero. Then
$\{\ket{\beta}: \beta\in V\}$ becomes a complete orthonormal basis
of $l^2(V)$. The adjacency matrix is considered as an operator
acting in $l^2(V)$ in such a way that
\begin{equation}
A\ket{\beta}=\sum_{\alpha\sim \beta}\ket{\alpha}.
\end{equation}
\subsection{stratification}
For $x\neq y$ let $\partial(x,y)$ be the length of the shortest walk
connecting $x$ and $y$. By definition $\partial(x,x)=0$ for all
$x\in V$. The graph becomes a metric space with the distance
function $\partial$. Note that $\partial(x,y)=1$ if and only if
$x\sim y$. We fix a point $o \in V$ as an origin of the graph,
called reference vertex. Then, the graph $\Gamma$ is stratified into
a disjoint union of strata:
\begin{equation}\label{strat}
V=\bigcup_{i=0}^{\infty}\Gamma_{i}(o),\;\ \Gamma_i(o):=\{\alpha\in
V: \partial(\alpha,o)=i\}
\end{equation}
Note that $\Gamma_i(o)=\emptyset$ may occur for some $i \geq 1$. In
that case we have $\Gamma_i(o)= \Gamma_{i+1}(o)=...= \emptyset$.
With each stratum $\Gamma_i(o)$ we associate a unit vector in
$l^2(V)$ defined by
\begin{equation}\label{unitv}
\ket{\phi_{i}}=\frac{1}{\sqrt{\kappa_i}}\sum_{\alpha\in
\Gamma_{i}(o)}\ket{\alpha},
\end{equation}
where, $\kappa_i=|\Gamma_{i}(o)|$ is called the $i$-th valency of
the graph ($\kappa_i:=|\{\gamma:\partial(o,\gamma)=i
\}|=|\Gamma_{i}(o)|$).
  The closed subspace of $l^2(V)$ spanned by
$\{\ket{\phi_{i}}\}$ is denoted by $\Lambda(\Gamma)$. Since
$\{\ket{\phi_{i}}\}$ becomes a complete orthonormal basis of
$\Lambda(\Gamma)$, we often write
\begin{equation}
\Lambda(\Gamma)=\sum_{i}\oplus \textbf{C}\ket{\phi_{i}}.
\end{equation}
\subsection{Pseudo-distance-regular graphs}
Given a vertex $\alpha\in V$ of a graph $\Gamma$, consider the
stratification (\ref{strat}) with respect to $\alpha$ such that
$\Gamma_i(\alpha)=\emptyset$ for $i > d$. Then, we say that
$\Gamma$ is pseudo-distance-regular around vertex $\alpha$
whenever for any $\beta \in \Gamma_k(\alpha)$ and $0 \leq k \leq
d$, the numbers
\begin{equation}\label{pseudo}
c_k(\beta):=\frac{1}{\kappa(\beta)}\sum_{\gamma\in
\Gamma_1(\beta)\cap \Gamma_{k-1}(\alpha)}\kappa(\gamma),\;\
a_k(\beta):=\frac{1}{\kappa(\beta)}\sum_{\gamma\in
\Gamma_1(\beta)\cap \Gamma_{k}(\alpha)}\kappa(\gamma),\;\
b_k(\beta):=\frac{1}{\kappa(\beta)}\sum_{\gamma\in
\Gamma_1(\beta)\cap \Gamma_{k+1}(\alpha)}\kappa(\gamma),
\end{equation}
do not depend on the considered vertex $\beta \in
\Gamma_k(\alpha)$, but only on the value of $k$. In such a case,
we denote them by $c_k$, $a_k$ and $b_k$ respectively. Then, the
matrix
\begin{equation}\label{intpseudo}
\left(\begin{array}{ccccc}
  0 & c_1 & \ldots & c_{d-1} & c_d \\
  a_0 & a_1 & \ldots & a_{d-1} & a_d \\
  b_0 & b_1 & \ldots & b_{d-1} & 0
\end{array}\right)
\end{equation}
is called the (pseudo-)intersection array around vertex $\alpha$
of $\Gamma$. It is shown in Ref. \cite{21} that this is a
generalization of the concept of distance-regularity around a
vertex (which in turn is a generalization of distance-regularity).
It should be noticed that for regular graphs $\Gamma$
($\kappa(\beta)=\kappa$ for all $\beta\in V$), the numbers
$c_k,a_k$ and $b_k$ read as
\begin{equation}\label{pseudo'}
c_k=|\Gamma_1(\beta)\cap \Gamma_{k-1}(\alpha)|,\;\
a_k=|\Gamma_1(\beta)\cap \Gamma_{k}(\alpha)|,\;\
b_k=|\Gamma_1(\beta)\cap \Gamma_{k+1}(\alpha)|,
\end{equation}
where we tacitly understand that $\Gamma_{-1}(\alpha)=\emptyset$
(see Fig. $1$). The intersection numbers (\ref{pseudo'}) and the
valencies $\kappa_i=|\Gamma_i(\alpha)|$ satisfy the following
obvious conditions
$$a_i+b_i+c_i=\kappa,\;\;\ \kappa_{i-1}b_{i-1}=\kappa_ic_i ,\;\;\
i=1,...,d,$$
\begin{equation}\label{intersec'}
\kappa_0=c_1=1,\;\;\;\ b_0=\kappa_1=\kappa, \;\;\;\ (c_0=b_d=0).
\end{equation}

One should notice that, the definition of pseudo-distance regular
graphs together with the Eq.(\ref{intersec'}), imply that in
general, the valencies $\kappa_i$ ( the size of the $i$-th
stratum) for $i=0,1,...,d$ do not depend on the considered vertex
$\beta \in \Gamma_k(\alpha)$, but only on the value of $k$.

The notion of pseudo-distance regularity has a close relation with
the concept of QD type graphs introduced by Obata \cite{obata},
such that for the adjacency matrices of this type of graphs, one
can obtain a quantum decomposition associated with the
stratification (\ref{strat}) as
\begin{equation}\label{QD'}
A=A^++A^-+A^0,
\end{equation}
where, the matrices $A^+$, $A^-$ and $A^0$ are defined as follows:
for $\beta\in \Gamma_k(\alpha)$, we set
$$
\bigl(A^{+})_{\beta\delta}\;=\;\cases{A_{\beta\delta} & if $\;
\delta\in \Gamma_{k+1}(\alpha)$ \cr 0 & otherwise\cr}, $$
$$
\bigl(A^{-})_{\beta\delta}\;=\;\cases{A_{\beta\delta} & if $\;
\delta\in \Gamma_{k-1}(\alpha)$ \cr 0 & otherwise\cr},
$$$$
\bigl(A^{0})_{\beta\delta}\;=\;\cases{A_{\beta\delta} & if $\;
\delta\in \Gamma_{k}(\alpha)$ \cr 0 & otherwise\cr},
$$
Or equivalently, for $\ket{k, \beta}$,
\begin{equation}\label{QD}
A^+\ket{k, \beta}=\sum_{\delta\in \Gamma_{k+1}(\alpha),\\
\delta\sim\beta}\ket{k+1, \delta},\;\ A^-\ket{k,
\beta}=\sum_{\delta\in
\Gamma_{k-1}(\alpha),\delta\sim\beta}\ket{k-1, \delta},,\;\
A^0\ket{k, \beta}=\sum_{\delta\in
\Gamma_{k}(\alpha),\delta\sim\beta}\ket{k, \delta},
\end{equation}
Since $\beta\in \Gamma_k(\alpha)$ and $\beta \sim \delta$ then
$\delta\in \Gamma_{k-1}(\alpha) \cup \Gamma_k(\alpha)\cup
\Gamma_{k+1}(\alpha)$.

It has been shown in Ref. \cite{obata} that, if $\Lambda(\Gamma)$
is invariant under the quantum components $A^+,A^-$ and $A^0$,
then there exist two sequences (called Szeg\"{o}- Jacobi
sequences) $\{\omega_l\}_{l=1}^{\infty}$ and
$\{\alpha_l\}_{l=1}^{\infty}$ derived from $A$ such that
$$A^+\ket{\phi_l}=\sqrt{\omega_{l+1}}\ket{\phi_{l+1}},\;\ l\geq0 \;\
,$$$$ A^-\ket{\phi_0}=0,\;\
A^-\ket{\phi_l}=\sqrt{\omega_l}\ket{\phi_{l-1}},\;\ l\geq 1,$$
\begin{equation}\label{QD1}
A^0\ket{\phi_l}=\alpha_{l}\ket{\phi_l}, l\geq0,
\end{equation}
where $\omega_{l+1} = \frac{\kappa_{l+1}}{\kappa_l}
\kappa^2_-(j)$, $\kappa_-(j) = |\{i\in \Gamma_l(\alpha) :\;\ i
\sim j\}|$ for $j\in \Gamma_{l+1}(\alpha)$ and $\alpha_{l}=
\kappa_0(j)$, such that $\kappa_0(j) =|\{i \in V_l; i \sim j\}|$
for $j\in \Gamma_l(\alpha)$, for $l\geq0$. One can easily check
that the coefficients $\alpha_i$ and $\omega_i$ are given by
\begin{equation}\label{omegal}
\alpha_k\equiv a_k=\kappa-b_{k}-c_{k},\;\;\;\;\
\omega_k\equiv\beta^2_k=b_{k-1}c_{k},\;\;\ k=1,...,d.
\end{equation}

By using (\ref{QD'}) and (\ref{QD1}), one can obtain
\begin{equation}\label{QD2}
A\ket{\phi_l}=\beta_{l+1}\ket{\phi_{l+1}}+\alpha_{l}\ket{\phi_l}+\beta_l\ket{\phi_{l-1}},\;\
l\geq0,
\end{equation}
with $\beta_l:=\sqrt{\omega_l}$. Then, by using (\ref{QD2}), one
can deduce that
\begin{equation}\label{dispoly0}
\ket{\phi_l}=P_l(A)\ket{\phi_0}, \;\;\ l=1,2,...,d ,
\end{equation}
where, $P_l(A)=a_0+a_1A+a_2A^2+...+a_{l}A^l$ is a polynomial of
degree $l$ in indeterminate $A$ and conversely $A^l$ can be
written as a linear combination of $P_0(A), P_1(A), ..., P_d(A)$.
In fact, it can be shown that \cite{jss}, the unit vectors
$\ket{\phi_l}$ for $l=0,1,...,d$ are the orthonormal basis
produced by application of the orthonormalization process (Lanczos
algorithm) to the so called Krylov basis
$\{\ket{\phi_0},A\ket{\phi_0},A^2\ket{\phi_0},...,A^{d}\ket{\phi_0}\}$
(for more details see for example \cite{js}, \cite{jss}).
\subsection{Spectral distribution associated with the network}
In this subsection we recall some facts about the spectral
distribution associated with the adjacency matrix of the network. In
fact, the spectral analysis of operators is an important issue in
quantum mechanics, operator theory and mathematical physics
\cite{simon, Hislop}. Since the advent of random matrix theory
(RMT), there has been considerable interest in the statistical
analysis of spectra \cite{rmt,rmt1,rmt3}. Also, the two-point
resistance has a probabilistic interpretation based on classical
random walker walking on the network. Indeed, the connection between
random walks and electric networks has been recognized for some time
(see e.g. \cite{Kakutani, Kemeny, Kelly} ), where one can establish
a connection between the electrical concepts of current and voltage
and corresponding descriptive quantities of random walks regarded as
finite state Markov chains (for more details see \cite{2}). Also, by
adapting the random-walk dynamics and mean-field theory it has been
studied that \cite{Bosiljka}, how the growth of a conducting
network, such as electrical or electronic circuits, interferes with
the current flow through the underlying evolving graphs. In Ref.
\cite{jss1}, it has been shown that, there is also connection
between the mathematical techniques introduced in previous
subsections and this subsection such as Hilbert space of the
stratification and spectral techniques (which have been employed in
\cite{js,jsannals,jss,rootlatt,laplac} for investigating continuous
time quantum walk on graphs), and electrical concept of resistance
between two arbitrary nodes of regular networks, and so the same
techniques can be used for calculating the resistance. Note that,
although we take the spectral approach to define the effective
resistance in terms of orthogonal polynomials (which are orthogonal
with respect to the spectral distribution $\mu$ associated with the
network) with three term recursion relations, in practice as it will
be seen in section $3$, the effective resistances will be calculated
without any need to evaluate the spectral distribution $\mu$.

It is well known that, for any pair $(A,\ket{\phi_0})$ of a matrix
$A$ and a vector $\ket{\phi_0}$, it can be assigned a measure
$\mu$ as follows
\begin{equation}\label{sp1}
\mu(x)=\braket{ \phi_0}{E(x)|\phi_0},
\end{equation}
 where
$E(x)=\sum_i|u_i\rangle\langle u_i|$ is the operator of projection
onto the eigenspace of $A$ corresponding to eigenvalue $x$, i.e.,
\begin{equation}
A=\int x E(x)dx.
\end{equation}
It is easy to see that, for any polynomial $P(A)$ we have
\begin{equation}\label{sp2}
P(A)=\int P(x)E(x)dx,
\end{equation}
where for discrete spectrum the above integrals are replaced by
summation. Therefore, using the relations (\ref{sp1}) and
(\ref{sp2}), the expectation value of powers of adjacency matrix
$A$ over starting site $\ket{\phi_0}$ can be written as
\begin{equation}\label{v2}
\braket{\phi_{0}}{A^m|\phi_0}=\int_{R}x^m\mu(dx), \;\;\;\;\
m=0,1,2,....
\end{equation}
The existence of a spectral distribution satisfying (\ref{v2}) is
a consequence of Hamburger's theorem, see e.g., Shohat and
Tamarkin [\cite{st}, Theorem 1.2].

Obviously relation (\ref{v2}) implies an isomorphism from the
Hilbert space of the stratification onto the closed linear span of
the orthogonal polynomials with respect to the measure $\mu$. More
clearly, the orthonormality of the unit vectors $\ket{\phi_i}$
implies that
\begin{equation}\label{ortpo}
\delta_{ij}=\langle\phi_i|\phi_j\rangle=\int_{R}P_i(x)P_j(x)\mu(dx),
\end{equation}
where, we have used the Eq. (\ref{dispoly0}) to write
$\ket{\phi_i}=P_i(A)\ket{\phi_0}$. Now, by substituting
(\ref{dispoly0}) in (\ref{QD2}), we get three term recursion
relations between polynomials $P_j(A)$, which leads to the following
three term recursion relations between polynomials $P_j(x)$
\begin{equation}\label{trt0}
xP_{k}(x)=\beta_{k+1}P_{k+1}(x)+\alpha_kP_{k}(x)+\beta_kP_{k-1}(x)
\end{equation}
for $k=0,...,d-1$, with $P_0(x)=1$. Multiplying two sides of
(\ref{trt0}) by $\beta_1...\beta_k$ we obtain
\begin{equation}
\beta_1...\beta_kxP_{k}(x)=\beta_1...\beta_{k+1}P_{k+1}(x)+\alpha_k\beta_1...\beta_kP_{k}(x)+\beta_k^2.\beta_1...\beta_{k-1}P_{k-1}(x).
\end{equation}
By rescaling $P_k$ as $Q_k=\beta_1...\beta_kP_k$, the spectral
distribution $\mu$ under question is characterized by the property
of orthonormal polynomials $\{Q_k\}$ defined recurrently by
$$ Q_0(x)=1, \;\;\;\;\;\
Q_1(x)=x,$$
\begin{equation}\label{op}
xQ_k(x)=Q_{k+1}(x)+\alpha_{k}Q_k(x)+\beta_k^2Q_{k-1}(x),\;\;\
k\geq 1
\end{equation}
(for more details see Refs.\cite{tsc, st, obh,obah}).

It should be noticed that, the starting values of the recurrence
(\ref{op}) are $Q_{-1}=0$, $Q_0=1$. If one starts from $q_{-1}=-1$,
$q_0=0$ and uses the same recurrence (with $\omega_0=1$)
\begin{equation}\label{op'}
xq_{n}(x)=q_{n+1}(x)+\alpha_nq_n(x)+\omega_nq_{n-1}(x),
\end{equation}
then $q_n$ is of degree $n-1$, and by Favard's theorem the different
$q_n$'s are orthogonal with respect to some measure. The $q_n$'s are
called orthogonal polynomials of the second kind (sometimes for
$Q_n$ we say that they are of the first kind). They can also be
written in the form
\begin{equation}\label{op''}
q_{n}(z)=\int_{R}\frac{Q_m(z)-Q_m(x)}{z-x}d\mu(x),
\end{equation}
(for more details see for example \cite{totik}).
\section{Two-point resistances in regular resistor networks}
A classic problem in electric circuit theory studied by numerous
authors over many years, is the computation of the resistance
between two nodes in a resistor network (see, e.g., \cite{Cserti}).
The results obtained in this section show that, there is a close
connection between the techniques introduced in section $2$ such as
Hilbert space of the stratification and the orthogonal polynomials
of the first and second kind associated with the networks and
electrical concept of resistance between two arbitrary nodes of the
networks.

For a given regular network $\Gamma$ with $v$ vertices and adjacency
matrix $A$, let $r_{ij}=r_{ji}$ be the resistance of the resistor
connecting vertices $i$ and $j$. Hence, the conductance is
$c_{ij}=r^{-1}_{ij}=c_{ji}$ so that $c_{ij}=0$ if there is no
resistor connecting $i$ and $j$. Denote the electric potential at
the $i$-th vertex by $V_i$ and the net current flowing into the
network at the $i$-th vertex by $I_i$ (which is zero if the $i$-th
vertex is not connected to the external world). Since there exist no
sinks or sources of current including the external world, we have
the constraint $\sum_{i=1}^vI_i=0$. The Kirchhoff law states
\begin{equation}\label{resistor}
\sum_{j=1,j\neq i}^vc_{ij}(V_i-V_j)=I_i,\;\;\  i=1,2,...,v.
\end{equation}
Explicitly, Eq.(\ref{resistor}) reads
\begin{equation}\label{resistor1}
L\vec{V}=\vec{I},
\end{equation}
where, $\vec{V}$ and $\vec{I}$ are $v$-vectors whose components
are $V_i$ and $I_i$, respectively and
\begin{equation}\label{laplas}
L=\sum_{i}c_i|i\rangle\langle i|-\sum_{i,j}c_{ij}|i\rangle\langle
j|
\end{equation}
is the Laplacian of the graph $\Gamma$ with
\begin{equation}
c_i\equiv \sum_{j=1,j\neq i}^vc_{ij},
\end{equation}
for each vertex $\alpha$. Hereafter, we will assume that all nonzero
resistances are equal to $1$, then the off-diagonal elements of $-L$
are precisely those of $A$, i.e.,
\begin{equation}\label{laplas1}
L=\kappa I-A,
\end{equation}
with $\kappa\equiv \kappa_1=deg(\alpha)$, for each vertex $\alpha$.
It should be noticed that, $L$ has eigenvector $(1,1,...,1)^t$ with
eigenvalue $0$. Therefore, $L$ is not invertible and so we define
the psudo-inverse of $L$ as
\begin{equation}\label{inv.laplas}
L^{-1}=\sum_{i,\lambda_i\neq0} {\lambda}^{-1}_iE_i,
\end{equation}
where, $E_i$ is the operator of projection onto the eigenspace of
$L^{-1}$ corresponding to eigenvalue $\lambda_i$. It has been shown
that, the two-point resistances $R_{\alpha\beta}$ are given by
\begin{equation}\label{eq.res.}
R_{\alpha\beta}=\langle \alpha|L^{-1}|\alpha\rangle+\langle
\beta|L^{-1}|\beta\rangle-\langle \alpha|L^{-1}|\beta\rangle-\langle
\beta|L^{-1}|\alpha\rangle.
\end{equation}
This formula may be formally derived using Kirchoff 's laws, and
seems to have been long known in the electrical engineering
literature, with it appearing in several texts, such as Ref.
\cite{12}.

Now, consider two nodes $\alpha,\beta\in V$ such that
$L^{-1}_{\alpha\alpha}=L^{-1}_{\beta\beta}$ (as we will see in
subsection $4.1$, for distance-regular graphs as resistor networks,
the diagonal entries of $L^{-1}$ are independent of the vertex,
i.e., for all $\alpha,\beta\in V$, we have
$L^{-1}_{\alpha\alpha}=L^{-1}_{\beta\beta}$). Then, from the
relation (\ref{eq.res.}) and the fact that $L^{-1}$ is a real
matrix, we can obtain the two-point resistance between the nodes
$\alpha$ and $\beta$ as follows
\begin{equation}\label{eq.res.dist.}
R_{\alpha\beta}=2(L^{-1}_{\alpha\alpha}-L^{-1}_{\alpha\beta}).
\end{equation}
It should be noticed that, due to the stratification of the network,
all of the nodes belonging to the same stratum with respect to the
reference node (a node which stratification is done with respect to
it), possess the same effective resistance with respect to the
reference node (the proof is given in the appendix). More clearly,
in order to evaluate the effective resistance between two nodes
$\alpha$ and $\beta$ of a network, we consider one of the nodes, say
$\alpha$, as reference node and stratify the network with respect to
$\alpha$. Then, $\beta$ will belong to one of the strata with
respect to $\alpha$, say the $m$-th stratum $\Gamma_m(\alpha)$. Now,
for all $\beta\in \Gamma_m(\alpha)$, one can write
\begin{equation}\label{eq.res.dist.1}
L^{-1}_{\alpha\beta^{(m)}}=\langle\alpha|L^{-1}|\beta\rangle=\frac{1}{\sqrt{\kappa_m}}\langle\alpha|L^{-1}|\phi_m\rangle=\frac{1}{\sqrt{\kappa_m}}\langle\alpha|P_m(A)L^{-1}|\alpha\rangle.
\end{equation}
where, we have used the equations (\ref{unitv}) and (\ref{dispoly0})
and the lemma given in the appendix. Then, by using
(\ref{eq.res.dist.}), we obtain for all $\beta\in \Gamma_m(\alpha)$
$$R_{\alpha\beta^{(m)}}=\frac{2}{\sqrt{\kappa_m}}\{\sqrt{\kappa_m}L^{-1}_{\alpha\alpha}-(P_m(A)L^{-1})_{\alpha\alpha}\}=\frac{2}{\sqrt{\kappa_m\omega_1...\omega_m}}\langle\alpha|\frac{\sqrt{\kappa_m\omega_1...\omega_m}1-Q_m(A)}{\kappa1-A}|\alpha\rangle=$$
\begin{equation}\label{eq.res.dist.2}
\frac{2}{\sqrt{\kappa_m\omega_1...\omega_m}}\int_{R-\{\kappa\}}\frac{Q_m(\kappa)-Q_m(x)}{\kappa-x}d\mu(x).
\end{equation}
where, the upper index $m$ in $L^{-1}_{\alpha\beta^{(m)}}$ and
$R_{\alpha\beta^{(m)}}$ indicate that $\beta$ belongs to the
$m$-th stratum with respect to $\alpha$. Note that, we have
substituted $P_m(x)=\frac{1}{\sqrt{\omega_1...\omega_m}}Q_m(x)$
and used the equality
$Q_m(\kappa)=\sqrt{\kappa_m\omega_1...\omega_m}$ which can be
verified easily. It should be noticed that, the result
(\ref{eq.res.dist.2}) can be written as
\begin{equation}\label{eq.res.dist.3}
R_{\alpha\beta^{(m)}}=\frac{2}{\sqrt{\kappa_m\omega_1...\omega_m}}\{\int_{R}\frac{Q_m(\kappa)-Q_m(x)}{\kappa-x}d\mu(x)-
\frac{1}{N}(\frac{\partial}{\partial x}Q_m(x))|_{x=\kappa}\}.
\end{equation}
Now, by using (\ref{op''}) and (\ref{eq.res.dist.3}), we obtain the
main result of the paper as follows
\begin{equation}\label{eq.res.dist.4}
R_{\alpha\beta^{(m)}}=\frac{2}{\sqrt{\kappa_m\omega_1...\omega_m}}\{q_m(\kappa)-
\frac{1}{N}(\frac{\partial}{\partial
x}Q_m(x))|_{x=\kappa}\},\;\;\;\ m=1,2,...,d.
\end{equation}
We recall that, the result (\ref{eq.res.dist.4}) can be used only
for evaluating the effective resistance between nodes such that the
network is symmetric with respect to them in the sense that,
stratification of the network with respect to these nodes produces
the same strata. Then for such nodes $\alpha$ and $\beta$, we will
have $L^{-1}_{\alpha\alpha}=L^{-1}_{\beta\beta}$.
\section{Examples}
\subsection{Evaluation of effective resistances on distance-regular networks}
First, we recall the definition of special kind of
pseudo-distance-regular networks called
distance-regular networks:\\
\textbf{Definition.} A pseudo-distance regular network
$\Gamma=(V,E)$ is called distance-regular with diameter $d$ if for
all $k\in\{0,1,...,d\}$, and $\alpha,\beta\in V$ with $\beta\in
\Gamma_k(\alpha)$, the numbers $c_k(\beta)$, $a_k(\beta)$ and
$b_k(\beta)$ defined in (\ref{pseudo}) depend only on $k$ but do not
depend on the choice of $\alpha$ and $\beta$.

It should be noticed that, in distance-regular networks, the
$i$-th adjacency matrix of the network $\Gamma=(V,R)$ is defined
by
\begin{equation}\label{adji.}
\bigl(A_i)_{\alpha, \beta}\;=\;\cases{1 & if $\; \partial(\alpha,
\beta)=i$ \cr 0 & \mbox{otherwise}\cr}.
\end{equation}
Then, from the definition of $A_i$, for the reference state
$\ket{\phi_0}$ ($\ket{\phi_0}=\ket{o}$, with $o\in V$ as reference
vertex), we have
\begin{equation}\label{Foc1}
A_i\ket{\phi_0}=\sum_{\beta\in \Gamma_{i}(o)}\ket{\beta}.
\end{equation}
Then by using (\ref{unitv}) and (\ref{Foc1}), we have
\begin{equation}\label{Foc2}
A_i\ket{\phi_0}=\sqrt{\kappa_i}\ket{\phi_i}.
\end{equation}
Also, it can be shown that, for adjacency matrices of a distance
regular graph, we have
$$
A_1A_i=b_{i-1}A_{i-1}+a_iA_i+c_{i+1}A_{i+1}, \;\;\;\ \mbox{for}
\;\ i=1,2,...,d-1,
$$
\begin{equation}\label{dra}
A_1A_d=b_{d-1}A_{d-1}+a_dA_d.
\end{equation}
Using the recursion relations (\ref{dra}), one can show that $A_i$
is a polynomial in $A_1$ of degree $i$, i.e., we have
\begin{equation}\label{dispoly}
A_i=P_i(A_1), \;\;\ i=1,2,...,d ,
\end{equation}
and conversely $A_1^i$ can be written as a linear combination of
$I, A_1, ..., A_d$ (for more details see for example \cite{js}).

Now, it should be noticed that, stratification of distance-regular
networks will be independent of the choice of the reference node
(the node which stratification is done with respect to it). Then,
clearly we will have $L^{-1}_{\alpha\alpha}=L^{-1}_{\beta\beta}$ for
all $\alpha,\beta\in V$ with $\alpha\neq \beta$ and consequently,
the result (\ref{eq.res.dist.4}) can be used for evaluation of the
effective resistance between any two arbitrary nodes $\alpha,\beta$;
It is sufficient to choose one of these nodes, say $\alpha$, as
reference node and stratify the network with respect to it. Then,
$\beta$ will be contained in one of the strata, say $m$-th stratum,
with respect to $\alpha$ and so the effective resistance between
$\alpha,\beta\in \Gamma_m(\alpha)$ can be
evaluated via Eq.(\ref{eq.res.dist.4}).\\
\textbf{1. Cycle network $C_{v}$}\\ A well known example of
distance-regular networks, is the cycle network $C_v$ with
$\kappa=2$. The network $C_v$ for $v=2k$ or $v=2k+1$ consists of
$k+1$ strata, where the intersection arrays for even and odd number
of vertices are given by
\begin{equation}\label{intcye.}
\{b_0,...,b_{k-1};c_1,...,c_k\}=\{2,1,...,1,1;1,...,1,2\}
\end{equation}
\begin{equation}\label{intcyo.}
\mbox{and}\;\
\{b_0,...,b_{k-1};c_1,...,c_k\}=\{2,1,...,1;1,...,1,1\},
\end{equation}
respectively. Then, by using (\ref{omegal}), for even $v=2k$ the QD
parameters are given by
\begin{equation}\label{QDcye.}
\alpha_i=0, \;\ i=0,1,...,k; \;\ \omega_1=\omega_k=2,\;\
\omega_i=1,\;\ i=2,...,k-1,
\end{equation}
where, for odd $v=2k+1$, we obtain
\begin{equation}\label{QDcyo.}
\alpha_i=0, \;\ i=0,1,...,k-1, \;\ \alpha_k=1;\;\ \omega_1=2,\;\
\omega_i=1,\;\ i=2,...,k.
\end{equation}
We consider $v=2k$ (the case $v=2k+1$ can be considered similarly).
Then, by using (\ref{op}) and (\ref{QDcye.}), one can obtain
\begin{equation}\label{Cypol}
Q_0=1,\;\ Q_1=x,\;\ Q_i(x)=2T_i(x/2),\;\ i>1;\;\
q_i(x)=Q_{i-1}(x),\;\ i\geq 1,
\end{equation}
where, $T_i$'s are Chebyshev polynomials in one variable, which are
recursively defined by
\begin{equation}
T_0=1,\;\ T_1=x,\;\ T_n(x)=2xT_{n-1}-T_{n-2},\;\ n>1.
\end{equation}
Then, by using (\ref{eq.res.dist.4}) and (\ref{Cypol}), the
effective resistance between any two nodes $\alpha,\beta\in
\Gamma_m(\alpha)$, is obtained as
$$R_{\alpha\beta^{(1)}}=q_1(2)-\frac{1}{2k}\frac{\partial}{\partial
x}Q_1(x)|_{x=2}=1-\frac{1}{2k}\;\ ,$$
\begin{equation}\label{Cyres}
R_{\alpha\beta^{(m)}}=2T_{m-1}(1)-\frac{1}{2k}T'_{m}(1),\;\;\;\
\mbox{for}\;\ m>1.
\end{equation}
\textbf{2. (d+1)-Simplex fractals}\\
$(d+1)$-simplex fractal is a generalization of a two dimensional
Sierpinski gasket to $d$-dimensions such that its subfractals are
$(d+1)$-simplices or $d$-dimensional polyhedra with
$S_(d+1)$-symmetry. In order to obtain a fractal with decimation
number $2$, we choose a $(d+1)$-simplex and divide all the links
(that is the lines connecting sites ) into $2$ parts and then draw
all possible $d$-dimensional hyperplanes through the links
parallel to the transverse $d$-simplices. Next, having omitted
every other innerpolyhedra, we repeat this process for the
remaining simplices or for the subfractals of next higher
generation. This way through $(d + 1)$-simplex fractals are
constructed. We label subfractals of generation $(n+1)$ in terms
of partition of $1$ into $(d + 1)$ positive integers $\lambda_1,
\lambda_2, ..., \lambda_{d+1}$. Each partition represents a
subfractal of generation $n$, and $\lambda$ shows the distance of
the corresponding subfractal from $d$-dimensional hyperplanes
which construct the $(d+1)$-simplex. On the other hand, each
vertex denoted by partition of $2$ into $(d + 1)$ non-negative
integers $\eta_1, \eta_2, · · · , \eta_{d+1}$ and obviously the
$i$-th vertex of subfractal $(\lambda_1, \lambda_2, ...,
\lambda_{d+1})$ is denoted by $\eta_j = \lambda_j + \delta_{i,j}$,
where $j=1,2,···,d + 1$ (for more details see
\cite{jrest},\cite{jrest1}).

Now, we consider the $(d+1)$-simplex fractal with decimation number
$b=2$ such that all of the $d+1$ vertices
$\underbrace{(20...0)}_{d+1}$, $\underbrace{(020...0)}_{d+1}$, ...,
$\underbrace{(0...02)}_{d+1}$ are connected via a resistor to each
other (see Fig. 2 (a) and Fig. 2 (b) for $d=2$ and $d=4$,
respectively). Then, the number of vertices of $(d+1)$-simplex
fractal is $N=C^{d+1}_1+C^{d+1}_2=\frac{(d+1)(d+2)}{2}$ such that
the degree of each vertex is $\kappa=2d$. Also, it can be easily
shown that the network has $3$ strata with respect to the reference
node $(200...0)$, where the unit vectors $\ket{\phi_i}$, $i=0,1,2$
are given as follows
$$\ket{\phi_0}=\underbrace{\ket{200...0}}_{d+1},$$
$$\ket{\phi_1}=\frac{1}{\sqrt{2d}}\sum_{i=2}^{d+1}(\ket{00...0\underbrace{2}_i0...0}+\ket{10...0\underbrace{1}_i0...0}),$$
\begin{equation}\label{simplex}
\ket{\phi_2}=\sqrt{\frac{2}{d(d-1)}}\sum_{i,j=2,...,d+1;
i<j}\ket{0...0\underbrace{1}_i0...0\underbrace{1}_j0...0}.
\end{equation}
Then, one can show that
$$A\ket{\phi_0}=\sqrt{2d}\ket{\phi_1},$$
$$A\ket{\phi_1}=\sqrt{2d}\ket{\phi_0}+d\ket{\phi_1}+2\sqrt{d-1}\ket{\phi_2},$$
\begin{equation}\label{simplex1}
A\ket{\phi_2}=2\sqrt{d-1}\ket{\phi_1}+2(d-2)\ket{\phi_2}.
\end{equation}
Also, by using the recursion relations (\ref{simplex1}) one can
easily show that the adjacency matrices $A\equiv A_1, A_2=J-A-I$
(where, $J$ is all one matrix) satisfy the following relations
$$A^2=2d. I_{\frac{(d+1)(d+2)}{2}}+d.A+4A_2,$$
\begin{equation}\label{simplex1'}
AA_2=(d-1)A+2(d-2)A_2.
\end{equation}

It should be noticed that if we stratify the network with respect
to another reference node such as $(1100...0)$, the unit vectors
will be obtained as
$$\ket{\phi_0}=\underbrace{\ket{1100...0}}_{d+1},$$
$$\ket{\phi_1}=\frac{1}{\sqrt{2d}}\{\sum_{i=3,...,d+1}(\ket{10...0\underbrace{1}_i0...0}+\ket{010...0\underbrace{1}_i0...0})+\ket{200...0}+\ket{020...0}\},$$
\begin{equation}\label{simplex5}
\ket{\phi_2}=\sqrt{\frac{2}{d(d-1)}}\{\sum_{i=3,...,d+1}\ket{00...0\underbrace{2}_i0...0}+\sum_{i,j=3,...,d+1;
i<j}\ket{00...0\underbrace{1}_i0...0\underbrace{1}_j0...0}\}.
\end{equation}
Then, one can show that, the same recursion relations as in
(\ref{simplex1}) and (\ref{simplex1'}) are satisfied for this
stratification, i.e., stratification of the network gives
three-term recursion relations independent of the choice of the
reference node and so the network is distance-regular.

Now, by using the equations (\ref{QD2}) and (\ref{simplex1}), one
can obtain
\begin{equation}\label{simplex2}
\alpha_0=0,\;\ \alpha_1=d,\;\ \alpha_2=2(d-2)\;\ ; \;\
\omega_1=2d,\;\ \omega_2=4(d-1).
\end{equation}
Then, by using the recursion relations (\ref{op}) and (\ref{op'}),
one can see that
\begin{equation}\label{simplex3}
Q_1(x)=x,\;\ Q_2(x)=x^2-dx-2d \;\ ; \;\ q_1(x)=1,\;\ q_2(x)=x-d.
\end{equation}
It should be noticed that, the intersection array of the $(d+1)$-
simplex fractal with decimation number $b=2$, can be evaluated by
using (\ref{omegal}) and (\ref{simplex2}) as:
$\{b_0,b_1;c_1,c_2\}=\{2d,d-1;1,4\}$ which implies that
$\kappa=b_0=2d, \kappa_2=\frac{\kappa b_1}{c_2}=\frac{d(d-1)}{2}$.
Now, by using (\ref{eq.res.dist.4}) and (\ref{simplex3}), the
effective resistance between any node $\alpha\in V$ and $\beta\in
\Gamma_{m}(\alpha)$ for $m=1,2$ is given by
$$
R_{\alpha,\beta^{(1)}}=\frac{2}{\kappa}(1-\frac{1}{N})=\frac{1}{d}(1-\frac{2}{(d+1)(d+2)})=\frac{d+3}{(d+1)(d+2)},$$
\begin{equation}\label{simplex4}
R_{\alpha,\beta^{(2)}}=\frac{1}{d(d-1)}\{q_2(2d)-\frac{1}{N}(2x-d)|_{x=2d}\}=\frac{1}{d(d-1)}\{d-\frac{2}{(d+1)(d+2)}.3d\}=\frac{d+4}{(d+1)(d+2)}.
\end{equation}
 \textbf{3. Hexagon network}\\
Consider the hexagon network with its diameters shown in Fig. $3$.
As the figure shows, this network has $6$ nodes with intersection
array $\{b_0,b_1;c_1,c_2\}=\{3,2;1,3\}$. Then by using
(\ref{intersec'}) and (\ref{omegal}), one can obtain
\begin{equation}\label{hexint}
\kappa=3,\;\ \kappa_2=2;\;\;\ \alpha_0=\alpha_1=\alpha_2=0,\;\
\omega_1=3,\;\ \omega_2=6.
\end{equation}
Then, by using (\ref{op}) and (\ref{hexint}), we obtain
\begin{equation}\label{hexpol}
Q_0=1,\;\ Q_1=x,\;\ Q_2(x)=x^2-3\;\ ;\;\ q_1(x)=1,\;\ q_2(x)=x.
\end{equation}
Now, by using (\ref{eq.res.dist.4}) and (\ref{hexpol}), the
effective resistance between any node $\alpha\in V$ and $\beta\in
\Gamma_{m}(\alpha)$ for $m=1,2$ is given by
\begin{equation}\label{hexres}
R_{\alpha,\beta^{(1)}}=\frac{2}{3}(1-\frac{1}{6})=\frac{5}{9},\;\;\
R_{\alpha,\beta^{(2)}}=\frac{1}{3}\{q_2(3)-\frac{1}{6}(2x)|_{x=3}\}=\frac{2}{3}.
\end{equation}
\textbf{4. A bipartite distance-regular network with $2n$ nodes}\\
Consider a distance-regular network with $2n$ nodes and adjacency
matrices as
\begin{equation}
A_1=\sigma_x\otimes (J_n-I_n),\;\ A_2=I_2\otimes(J_n-I_n),\;\;\
A_3=\sigma_x\otimes I_n,
\end{equation}
where, $J_n$ is an $n\times n$ matrix such that all of its entries
are one. Then, one can show that
\begin{equation}\label{Eqx}
A^2_1=(n-1)I_{2n}+(n-2)A_2,\;\ A_1A_2=(n-2)A_1+(n-1)A_3,\;\;\
A_1A_3=A_2,
\end{equation}
By using (\ref{dra}) and (\ref{Eqx}), the intersection array of
the network is given by
$$\{b_0,b_1,b_2;c_1,c_2,c_3\}=\{n-1,n-2,1;1,n-2,n-1\}.$$
Then, by using (\ref{intersec'}) and (\ref{omegal}), one can
obtain
$$\kappa=n-1,\;\ \kappa_2=n-1,\;\ \kappa_3=1,$$
\begin{equation}\label{Eqx1}
\alpha_i=0,\;\ i=0,1,2,\;\ ; \;\ \omega_1=n-1,\;\
\omega_2=(n-2)^2,\;\ \omega_3=n-1.
\end{equation}
Now, by using (\ref{op}) and (\ref{Eqx1}), we obtain
$$
Q_0=1,\;\ Q_1=x,\;\ Q_2(x)=x^2-n+1,\;\ Q_3(x)=x^3-(n^2-3n+3)x ;$$
\begin{equation}\label{Eqx2}
q_1(x)=1,\;\ q_2(x)=x,\;\ q_3(x)=x^2-(n-2)^2.
\end{equation}
So, by using (\ref{eq.res.dist.4}) and (\ref{Eqx2}), the effective
resistance between any node $\alpha\in V$ and $\beta\in
\Gamma_{m}(\alpha)$ for $m=1,2,3$ is given by
$$R_{\alpha,\beta^{(1)}}=\frac{2}{n-1}(1-\frac{1}{2n})=\frac{2n-1}{n(n-1)},$$
$$
R_{\alpha,\beta^{(2)}}=\frac{2}{(n-1)(n-2)}\{q_2(n-1)-\frac{1}{2n}(2x)|_{x=n-1}\}=\frac{2(n-1)}{n(n-2)},$$
\begin{equation}\label{hexres}
R_{\alpha,\beta^{(3)}}=\frac{2}{(n-1)(n-2)}\{q_3(n-1)-\frac{1}{2n}[3x^2-(n^2-3n+3)]|_{x=n-1}\}=\frac{2n-3}{(n-1)(n-2)}.
\end{equation}
\subsection{Evaluation of effective resistance
in examples of pseudo-distance-regular networks}
\textbf{1. Pseudo-distance-regular network derived from Hadamard network with $16$ nodes} \\
Consider the pseudo-distance-regular network shown in Fig. $4$. This
network is obtained from the Hadamard network with intersection
array $\{4,3,2,1;1,2,3,4\}$. As Fig. $4$ shows, the network is
symmetric with respect to the initial and final (horizontal) nodes
$1$ and $16\in \Gamma_4(1)$ and also with respect to the initial and
final (vertical) nodes $6$ and $11\in \Gamma_4(6)$. One should
notice that stratification of the network with respect to the nodes
$1$ and $16$ produces the same strata. For stratification with
respect to the node $1$ or $16$, we have
$$\kappa=4,\;\ \kappa_2=6,\;\ \kappa_3=4,\;\ \kappa_4=1,$$
$$\alpha_i=0,\;\ i=0,...,3;\;\;\ \omega_1=4,\;\ \omega_2=6,\;\ \omega_3=6,\;\ \omega_4=4.$$
(see Eq. (\ref{QD1})). Then, by using the recursion relations
(\ref{op}) and (\ref{op'}), one can obtain
\begin{equation}\label{had}
Q_4(x)=x^4-16x^2+24,\;\;\ q_4(x)=x^3-12x.
\end{equation}
Now, from Eq. (\ref{eq.res.dist.4}), the effective resistance
between nodes $1$ and $16\in \Gamma_4(1)$ is given by
\begin{equation}\label{had1}
R_{_{1,16}}=\frac{1}{12}\{q_4(4)-\frac{1}{16}(4x^3-32x)|_{x=4}\}=\frac{2}{3}.
\end{equation}
In order to evaluate the effective resistance between vertical nodes
$6$ and $11$, one must consider the stratification of the network
with respect to the node $6$ or $11$. For this stratification, we
obtain
$$\kappa=4,\;\ \kappa_2=4,\;\ \kappa_3=4,\;\ \kappa_4=3.$$
$$\alpha_i=0,\;\ i=0,...,3;\;\;\ \omega_1=4,\;\ \omega_2=4,\;\ \omega_3=1,\;\ \omega_4=12.$$
Then, by using the recursion relations (\ref{op}) and (\ref{op'}),
one can obtain
\begin{equation}\label{had}
Q_4(x)=x^4-9x^2+4,\;\;\ q_4(x)=x^3-5x.
\end{equation}
Therefore, from Eq. (\ref{eq.res.dist.4}), the effective resistance
between nodes $6$ and $11\in \Gamma_4(6)$ is given by
\begin{equation}\label{had1}
R_{_{6,11}}=\frac{1}{12}\{q_4(4)-\frac{1}{16}(4x^3-18x)|_{x=4}\}=\frac{65}{24}.
\end{equation}
\textbf{2. Pseudo-distance-regular network derived from Desargues}\\
This network has $20$ nodes with
$$\{b_0,b_1,b_2,b_3,b_4;c_1,c_2,c_3,c_4,c_5\}=\{3,2,2,1,1;1,1,2,2,3\}.$$
Then, by using (\ref{intersec'}) and (\ref{omegal}), one can obtain
$$\kappa=3,\;\ \kappa_2=6,\;\ \kappa_3=6, \;\ \kappa_4=3,\;\ \kappa_5=1$$
$$\alpha_i=0,\;\ i=0,1,...,5;\;\ \omega_1=3,\;\ \omega_2=2,\;\ \omega_3=4,\;\ \omega_4=2,\;\ \omega_5=3.$$
The stratification with respect to the initial node $1$ and the
final node $20$ produces the same strata. Therefore, the effective
resistance between the nodes $1$ and $20$ can be evaluated by using
the Eq. (\ref{eq.res.dist.4}). By using the recursion relations
(\ref{op}) and (\ref{op'}), we obtain
\begin{equation}\label{had}
Q_5(x)=x^5-11x^3+22x,\;\;\ q_5(x)=x^4-8x^2+4.
\end{equation}
Then, from Eq. (\ref{eq.res.dist.4}), the effective resistance
between nodes $1$ and $20\in \Gamma_5(1)$ is given by
\begin{equation}\label{had1}
R_{_{1,20}}=\frac{1}{6}\{q_5(3)-\frac{1}{20}(5x^4-33x^2+22)|_{x=3}\}=\frac{13}{12}.
\end{equation}
\textbf{3. Pseudo-distance-regular network derived from Hadamard network with $32$ nodes} \\
Consider the pseudo-distance-regular network obtained from the
Hadamard network with intersection array $\{8,7,4,1;1,4,7,8\}$ such
that, the network is symmetric with respect to the initial and final
(horizontal) nodes $1$ and $32\in \Gamma_4(1)$ and also with respect
to the initial and final (vertical) nodes $10$ and $23\in
\Gamma_4(10)$. One should notice that stratification of the network
with respect to the nodes $1$ and $32$ produces the same strata. For
stratification with respect to the node $1$ or $32$, we have
$$\kappa=8,\;\ \kappa_2=14,\;\ \kappa_3=8,\;\ \kappa_4=1,$$
$$\alpha_i=0,\;\ i=0,...,4;\;\;\ \omega_1=8,\;\ \omega_2=28,\;\ \omega_3=28,\;\ \omega_4=8.$$
(see Eq. (\ref{QD1})). Then, by using the recursion relations
(\ref{op}) and (\ref{op'}), one can obtain
\begin{equation}\label{had}
Q_4(x)=x^4-64x^2+224,\;\;\ q_4(x)=x^3-56x.
\end{equation}
Now, from Eq. (\ref{eq.res.dist.4}), the effective resistance
between nodes $1$ and $32\in \Gamma_4(1)$ is given by
\begin{equation}\label{had1}
R_{_{1,32}}=\frac{1}{112}\{q_4(8)-\frac{1}{32}(4x^3-128x)|_{x=8}\}=\frac{2}{7}.
\end{equation}
In order to evaluate the effective resistance between vertical nodes
$10$ and $23\in \Gamma_4(10)$, one must consider the stratification
of the network with respect to the node $10$ or $23$ and evaluate
$R_{_{10,23}}$ as before.\\
\textbf{5. A network with $9$ nodes} \\
Consider the network given in Fig. $5$ with $9$ nodes and the
following intersection array
$$\{b_0,b_1;c_1,c_2\}=\{4,2;1,2\}.$$
Then by using (\ref{intersec'}) and (\ref{omegal}), one can obtain
$$\kappa=4,\;\ \kappa_2=4, \;\ ; \;\ \alpha_0=0,\;\ \alpha_1=1,\;\ \alpha_2=2,\;\;\ \omega_1=\omega_2=4  .$$
As it can be seen from Fig. $7$, the stratification with respect to
the nodes $1$ and $9\in \Gamma_2(1)$ produces the same strata. Then,
by using the recursion relations (\ref{op}) and (\ref{op'}), one can
obtain
\begin{equation}\label{had}
Q_2(x)=x^2-x-4,\;\;\ q_2(x)=x-1.
\end{equation}
Now, from Eq. (\ref{eq.res.dist.4}), the effective resistance
between nodes $1$ and $9\in \Gamma_2(1)$ is given by
\begin{equation}\label{had1}
R_{_{1,9}}=\frac{1}{4}\{q_2(4)-\frac{1}{9}(2x-1)|_{x=4}\}=\frac{5}{9}.
\end{equation}
\textbf{6. A network with $16$ nodes} \\
Consider the network given in Fig. $6$ with $16$ nodes and the
following intersection array
$$\{b_0,b_1,b_2,b_3;c_1,c_2,c_3,c_4\}=\{4,3,2,1;1,2,3,4\}.$$
Then by using (\ref{intersec'}) and (\ref{omegal}), one can obtain
$$\kappa=4,\;\ \kappa_2=6, \;\ \kappa_3=4,\;\ \kappa_4=1 ; \;\ \alpha_i=0,\;\ i=0,1,...,4 ,\;\;\ \omega_1=4,\;\ \omega_2=\omega_3=6,\;\ \omega_4=4.$$
As it can be seen from Fig. $8$, the stratification with respect to
the nodes $1$ and $16\in \Gamma_2(1)$ produces the same strata.
Then, by using the recursion relations (\ref{op}) and (\ref{op'}),
one can obtain
\begin{equation}\label{had}
Q_2(x)=x^2-4,\;\;\ q_2(x)=x.
\end{equation}
Now, from Eq. (\ref{eq.res.dist.4}), the effective resistance
between nodes $1$ and $16\in \Gamma_2(1)$ is given by
\begin{equation}\label{had1}
R_{_{1,16}}=\frac{1}{6}\{q_2(4)-\frac{1}{16}(2x)|_{x=4}\}=\frac{7}{12}.
\end{equation}
\textbf{7. Generalized $G_2(m)$ type network} \\
Consider the network with $2(2^{m+1}-1)$ nodes and the following
intersection array
$$\{b_0,b_1,...,b_{m};c_1,c_2,...c_m,c_{m+1}\}=\{3,2,...,2;1,1,...1,3\}.$$
Then by using (\ref{intersec'}) and (\ref{omegal}), one can obtain
$$\kappa=3,\;\ \kappa_i=3.2^{i-1},\;\ i=2,...,m, \;\ \kappa_{m+1}=2^m,$$
$$\alpha_i=0,\;\ i=0,1,...,m+1;\;\ \omega_1=3,\;\ \omega_2=...=\omega_m=2,\;\  \omega_{m+1}=6.$$
Then, the stratification with respect to the initial node $1$ and
the final node $2^{m+2}-2\in \Gamma_1(1)$ produces the same strata.
Therefore, from the fact that $Q_1(x)=x$ and $q_1(x)=1$, the
effective resistance between nodes $1$ and $2^{m+2}-2\in
\Gamma_1(1)$ is obtained as
\begin{equation}\label{had1}
R_{_{1,2^{m+2}-2}}=\frac{2}{3}(1-\frac{1}{2(2^{m+1}-1)})=\frac{2^{m+2}-3}{3(2^{m+1}-1)}.
\end{equation}
\section{Conclusion}
Based on the stratification of the pseudo-distance-regular
networks and using spectral techniques, the evaluation of the
effective resistances on these networks was discussed. It was
shown that, in these types of networks, the effective resistances
between a node $\alpha$ and all nodes $\beta$ belonging to the
same stratum with respect to the $\alpha$ are the same. Then, an
explicit analytical formula for the effective resistance between
two nodes $\alpha,\beta$ of a pseudo-distance-regular resistor
network such that $L^{-1}_{\alpha\alpha}=L^{-1}_{\beta\beta}$
($L^{-1}$ is the pseudo-inverse of the Laplacian matrix of the
network) was given in terms of the first and second orthogonal
polynomials associated with the network. It was deduced that, the
obtained result can be used for evaluation of the effective
resistance between any two arbitrary nodes $\alpha,\beta$ in
distance-regular networks, where we have
$L^{-1}_{\alpha\alpha}=L^{-1}_{\beta\beta}$ for all nodes
$\alpha,\beta$.
\newpage
 \vspace{1cm}\setcounter{section}{0}
 \setcounter{equation}{0}
 \renewcommand{\theequation}{A-\roman{equation}}
  {\Large{Appendix}}\\
In this appendix we prove the following lemma in connection with
the equality of effective resistance between the reference node
and all of the nodes belonging to the same stratum with respect to
the reference node.

\textbf{Lemma.} Let $R_{\alpha\beta}$ denote the effective
resistance between nodes $\alpha,\beta\in V$. Then for
pseudo-distance-regular networks, by choosing one of the nodes,
say $\alpha$ as reference node, the effective resistance
$R_{\alpha\beta}$ is the same for all nodes $\beta\in
\Gamma_m(\alpha)$, where $m\in \{1,2,...,d\}$.\\
proof.\\
In order to prove the above lemma, we prove that in general for
any pseudo-distance-regular network with diameter $d$, we have
$\langle\phi_0|f(A)|\beta\rangle$ ($|\phi_0\rangle\equiv
|\alpha\rangle$) is the same for all $\beta\in \Gamma_l(\alpha)$,
i.e.,
\begin{equation}
\langle\phi_0|f(A)|\beta\rangle=\frac{1}{\sqrt{\kappa_l}}\langle\phi_0|f(A)|\phi_l\rangle,
\end{equation}
where, $f(A)$ is any function of the adjacency matrix $A$ of the
network such that $f(A)=\sum_{l=0}^da_kA^k$ and
\begin{equation}
|\phi_l\rangle=\frac{1}{\sqrt{\kappa_l}}\sum_{j\in
\Gamma_l(\alpha)}|j\rangle,\;\;\ l=0,1,...,d.
\end{equation}
To this aim, we take the Fourier transform of the unit vectors
$|\phi_l\rangle$ for $l=0,1,...,d$ as follows
\begin{equation}
|\phi_{l,k}\rangle=\frac{1}{\sqrt{\kappa_l}}\sum_{j\in
\Gamma_l(\alpha)}e^{2\pi i jk/\kappa_l}|j\rangle,\;\;\
l=0,1,...,d; \;\ k=0,1,...,\kappa_l-1.
\end{equation}
Now, we show that
$$\langle\phi_0|f(A)|\phi_{l,0}\rangle\neq 0,$$
\begin{equation}
\langle\phi_0|f(A)|\phi_{l,k}\rangle=0,\;\;\ \forall \;\
k=1,...,\kappa_l-1.
\end{equation}
To do so, we use the fact that there is a correspondence between
the basis $I,A,...,A^{d-1}$ and the orthogonal polynomials
$P_i(A)$ defined by Eq.(\ref{dispoly}). In fact, as it was
regarded previously (see arguments about the Eq.(\ref{dispoly}) ),
$A^l$ for $l=0,1,...,d$ can be written as a linear combination of
$P_0(A), P_1(A), ..., P_d(A)$.

It should be noticed that, in Krylov subspace projection methods,
approximations to the desired eigenpairs of an $n\times n$ matrix
$A$ are extracted from a $d$-dimensional Krylov subspace
\begin{equation}
K_d(\ket{\phi_0},A) = span\{\ket{\phi_0},A\ket{\phi_0},
\cdots,A^{d-1}\ket{\phi_0}\},
\end{equation}
where, $\ket{\phi_0}$ is often a randomly chosen starting vector
called reference state and $d \ll n$, i.e., the vectors
$\ket{\phi_0}, A\ket{\phi_0}, ... , A^{d-1}\ket{\phi_0}$
constitute a basis for the  Krylov subspace $
K_d(\ket{\phi_0},A)$. Then, the application of the
orthonormalization process (the Lanczos algorithm which is a
modified version of the classical Gram-Schmidt orthogonalization
process) to the Krylov basis $\{A^k\ket{\phi_0}\}_{k=0}^{d-1}$ is
equivalent to the construction of a sequence of orthonormal basis
$\ket{\phi_j}=P_j(A)\ket{\phi_0}$, where
$P_j(A)=a_0+a_1A+...+a_jA^j$  is a polynomial of degree $j$ in
indeterminate $A$.

As regards the above arguments, any function $f(A)$ can be
expanded as a linear combination of the polynomials $P_j(A)$,
i.e.,
\begin{equation}
f(A)=\sum_{j=0}^{d}b_j P_j(A).
\end{equation}
Then, we have
\begin{equation}\label{a1}
\langle\phi_0|f(A)|\phi_{l,k}\rangle=\sum_{j=0}^{d}b_j\langle\phi_0|
P_j(A)|\phi_{l,k}\rangle=\sum_{j=0}^{d}b_j\underbrace{\langle\phi_{j,0}|\phi_{l,k}\rangle}_{\delta_{jl}\delta_{k0}}=0,\;\;\
\forall \;\ k=1,...,\kappa_l-1.
\end{equation}

Now, let we denote $\langle\phi_0|f(A)|j\rangle$ by $x_{j}$ for
$j\in \Gamma_l(\alpha)$. Then, from (\ref{a1}) we have
\begin{equation}\label{a2}
F\left(\begin{array}{c}
  x_1 \\
  x_2 \\
  \vdots \\
  x_{\kappa_l} \\
\end{array}\right)=\left(\begin{array}{c}
  \langle\phi_0|f(A)|\phi_{l,0}\rangle \\
  0 \\
  \vdots \\
  0 \\
\end{array}\right).
\end{equation}
where, $F$ is the $\kappa_l\times \kappa_l$ discrete Fourier
transformation matrix(DFT) defined as
$F_{jk}=\frac{1}{\sqrt{\kappa_l}}e^{2\pi i jk/\kappa_l}$.
Therefore, by inverting $F$ in (\ref{a2}), we obtain
\begin{equation}\label{a3}
\left(\begin{array}{c}
  x_1 \\
  x_2 \\
  \vdots \\
  x_{\kappa_l} \\
\end{array}\right)=F^{\dag}\left(\begin{array}{c}
  \langle\phi_0|f(A)|\phi_{l,0}\rangle \\
  0 \\
  \vdots \\
  0 \\
\end{array}\right)=\frac{1}{\sqrt{\kappa_l}}\left(\begin{array}{c}
  \langle\phi_0|f(A)|\phi_{l,0}\rangle \\
  \langle\phi_0|f(A)|\phi_{l,0}\rangle\\
  \vdots \\
  \langle\phi_0|f(A)|\phi_{l,0}\rangle \\
\end{array}\right).
\end{equation}
That is, we obtain
$x_j=\langle\phi_0|f(A)|j\rangle=\frac{1}{\sqrt{\kappa_l}}\langle\phi_0|f(A)|\phi_{l,0}\rangle$
for all $j\in \Gamma_l(\alpha)$.

\newpage
{\bf Figure Captions}

{\bf Figure-1:} Shows edges through $\alpha$ and $\beta$ in a
pseudo-distance-regular graph.

{\bf Figure-2:} (a) Shows $3$-simplex fractal with decimation
number $b=2$ (b)Shows $5$-simplex fractal with decimation number
$b=2$.

{\bf Figure-3:} Shows the Hexagon network.

{\bf Figure-4:} Shows a pseudo-distance-regular network with $16$
nodes derived from Hadamard graph.

{\bf Figure-5:} Shows a pseudo-distance-regular network with $9$
nodes.

{\bf Figure-6:} Shows a pseudo-distance-regular network with $16$
nodes.

\begin{thebibliography}{99}
\bibitem{jss1}
M. A. Jafarizadeh, R. Sufiani and S. Jafarizadeh, J. Phys. A: Math.
Theor. 40 (2007) 4949-4972.
\bibitem{res2}
M. A. Jafarizadeh, R. Sufiani and S. Jafarizadeh, arXiv: 0705.2480
(2007).
\bibitem{21}
M.A. Fiol, E. Garriga, and J.L.A. Yebra, Locally
pseudo-distance-regular graphs, J. Combin. Theory Ser. B 68 (1996)
179-205.
\bibitem{obata}
N. Obata, Quantum Probabilistic Approach to Spectral Analysis of
Star Graphs, Interdisciplinary Information Sciences, Vol. 10 (2004)
41-52.
\bibitem{8}
G. Kirchhoff, Phys. Chem. 72 (1847) 497-508.
\bibitem{js}
M. A. Jafarizadeh and S. Salimi, J. Phys. A 39 (2006) 1-29 .
\bibitem{jsannals} M. A. Jafarizadeh and S. Salimi, Annals of
physics 322 (2007) 1005-1033.
\bibitem{tsc}
T. S. Chihara , {\it An Introduction to Orthogonal Polynomials},
Gordon and Breach, Science Publishers Inc (1978).
\bibitem{jss}
M. A. Jafarizadeh, S. Salimi and R. Sufiani, arXiv: quan-ph/0606241.
\bibitem{simon}
H. Cycon, R. Forese, W.Kirsch and B.Simon {\it Schrodinger
operators} (Springer-Verlag, 1987).
\bibitem{Hislop}
P. D. Hislop and I. M. Sigal, {\it Introduction to spectral theory:
With applications to schrodinger operators} (1995).
\bibitem{rmt}
C. E. Porter, {\it Statistical theories of spectra: fluctuations}
(Academic Press, New York, 1965).
\bibitem{rmt1}
 M. L. Mehta, {\it Random matrices},
2nd ed. (Academic Press, New York, 1991).
\bibitem{rmt3}
T. Guhr, A. Muller-Groeling, and H. A. Weidenm\"{u}ller, Phys. Rep.
299 (1998) 190.
\bibitem{Kakutani}
S. Kakutani, Proc. Jap. Acad., 21:227 (1945).
\bibitem{Kemeny}
J. G. Kemeny, J. L. Snell, and A.W. Knapp, {\it Denumerable Markov
Chains} (1966).
\bibitem{Kelly}
F. Kelly, {\it Reversibility and Stochastic Networks} (1979).
\bibitem{2}
P. G. Doyle and J. L. Snell,  Random Walks and Electric Networks
(The Carus MathematicalMonograph series 22) (Washington, DC: The
Mathematical Association of America)) pp 83–149 (Preprint
math.PR/0001057)(1984).
\bibitem{Bosiljka}
B. Tadic and V. Priezzhev, arXiv: cond-mat/0207100.
\bibitem{rootlatt}
M. A. Jafarizadeh and R. Sufiani, Physica A 381 (2007) 116-142.
\bibitem{laplac}
M. A. Jafarizadeh and R. Sufiani, arXiv: quant-ph/0704.2602 to be
published in International Journal of Quantum Information.
\bibitem{st}
J. A. Shohat, and J. D. Tamarkin, {\it The Problem of Moments,
American Mathematical
 Society}, Providence, RI (1943).
\bibitem{obh}
A. Hora, and N. Obata, Fundamental Problems in Quantum Physics,
World Scientific \textbf{284} (2003).
\bibitem{obah}
A. Hora, and N. Obata,  Quantum Information V, World Scientific,
Singapore (2002).
\bibitem{totik}
V. Totik, Surveys in Approximation Theory, Vol. 1 (2005) 70-125.
\bibitem{Cserti}
J. Cserti, Am. J. Phys. 68 (2000) 896  (Preprint cond-mat/9909120).
\bibitem{12}
S. Seshu and M. B. Reed, {\it Linear Graphs and Electrical
Networks}, Addison-Wesley, Reading, Massachusetts, (1961).
\bibitem{jrest}
M. A. Jafarizadeh, Physica A 287 (2000) 1-25.
\bibitem{jrest1}
M. A. Jafarizadeha, Europian Physical journal B, vol. 4 (1998) 103.
\end{thebibliography}
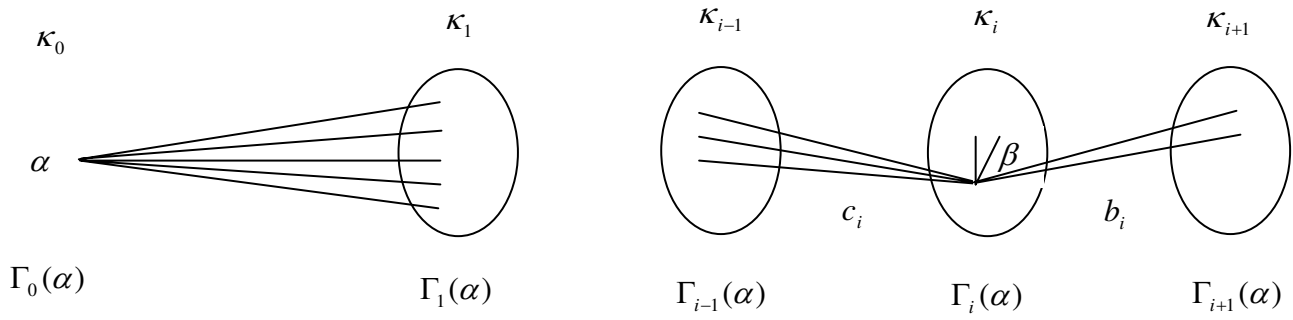
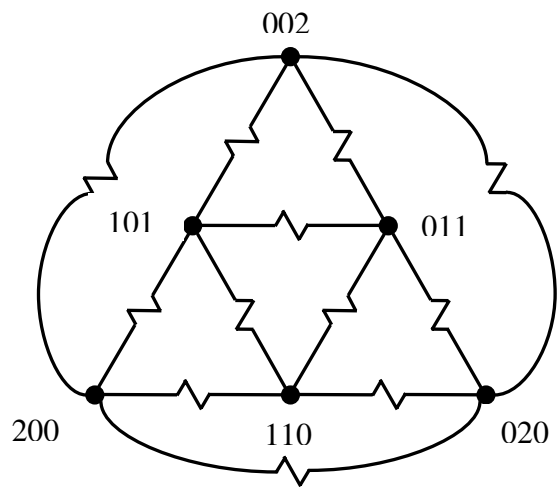
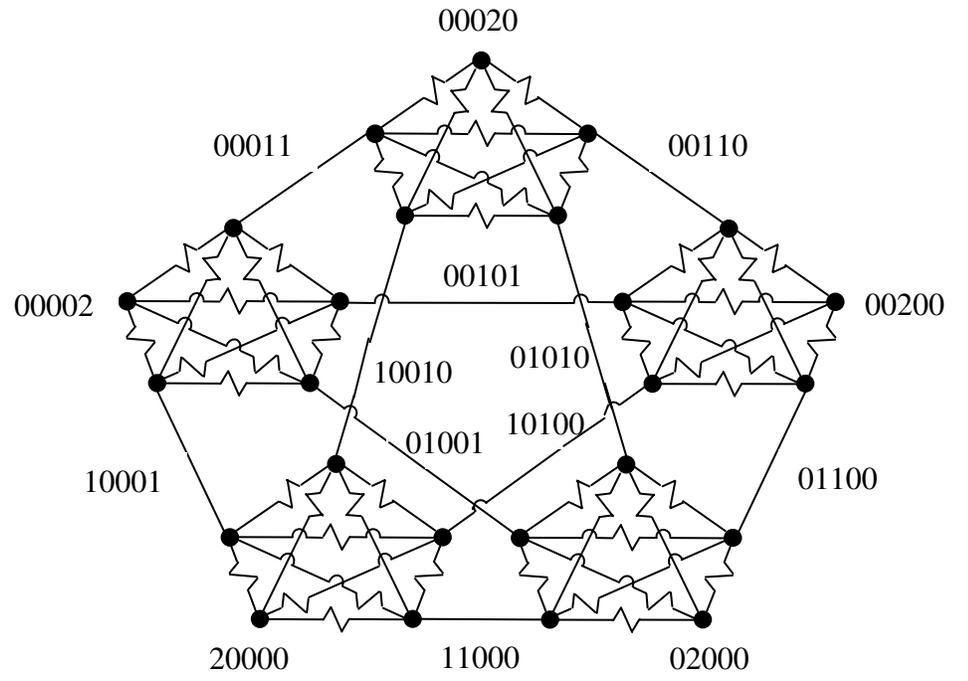
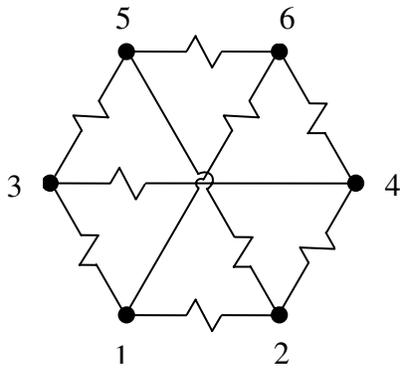
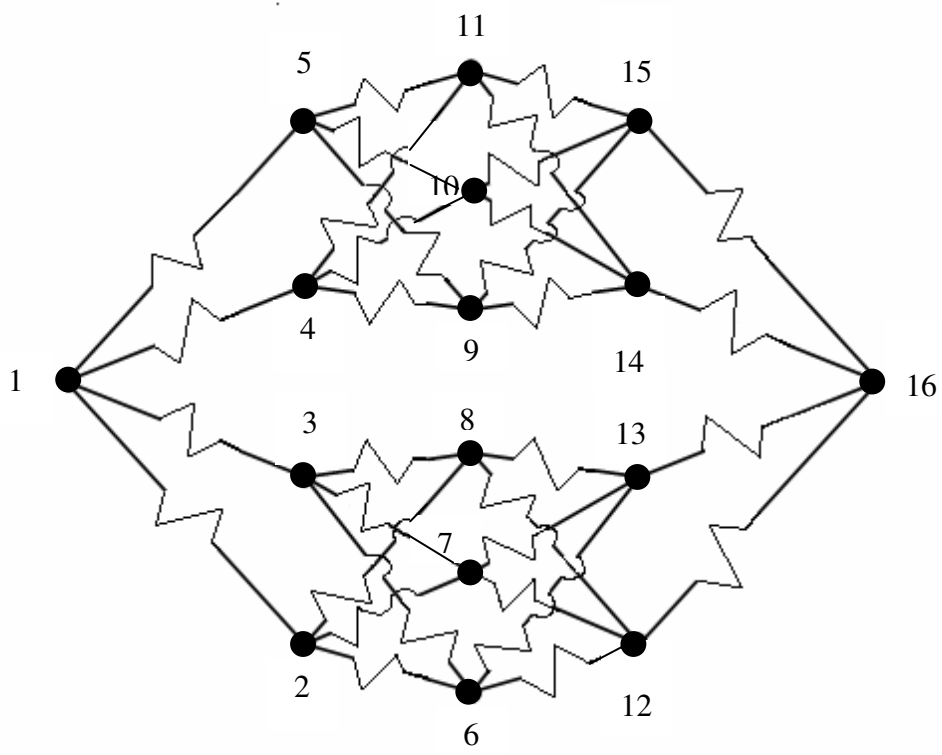
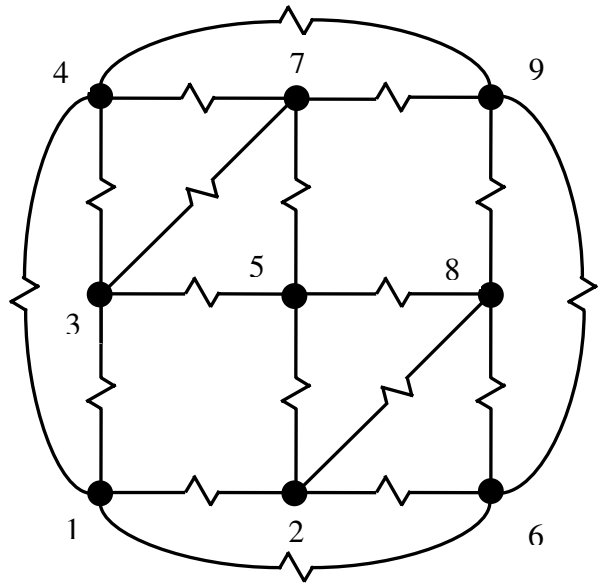
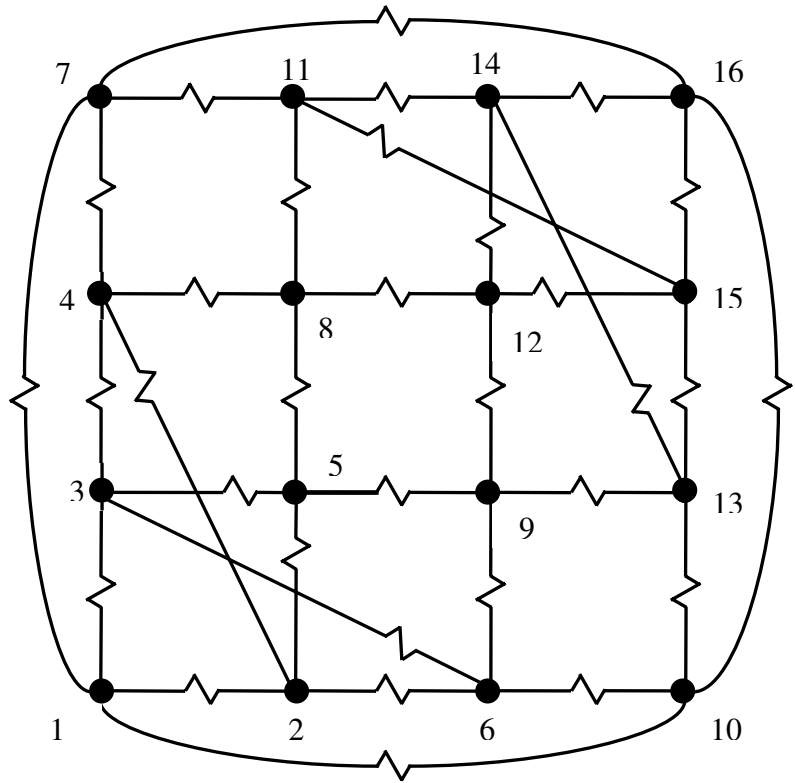
\end{document}